\documentclass[twoside,leqno,twocolumn]{article}

\usepackage{ltexpprt}
\usepackage{algorithm}
\usepackage[ruled,vlined,algo2e]{algorithm2e}
\usepackage[numbers,sort&compress]{natbib}
\usepackage{todonotes}
\usepackage{float}
\usepackage{amsfonts}
\usepackage{caption}
\SetKwInput{KwGiven}{Given}
\SetKwInput{KwFind}{Find}
\SetKwInput{KwWhere}{Where}
\SetKwInput{KwMin}{Minimizing}

\begin{document}

\title{\Large A General Framework For Task-Oriented Network Inference}
\author{Ivan Brugere\thanks{University of Illinois at Chicago} \\
\and
Chris Kanich\footnotemark[1]
\and
Tanya Y. Berger-Wolf\footnotemark[1]}
\date{}

\maketitle

%\setcounter{chapter}{2} % If you are doing your chapter as chapter one,
%\setcounter{section}{3} % comment these two lines out.

%\pagenumbering{arabic}
%\setcounter{page}{1}%Leave this line commented out.

\begin{abstract} \small\baselineskip=9pt We present a brief introduction to a flexible, general network inference framework which models data as a network space, sampled to optimize network structure to a particular \emph{task}. We introduce a formal problem statement related to influence maximization in networks, where the network structure is not given as input, but learned jointly with an influence maximization solution. \end{abstract}

\section{Introduction}

Networks are extensively studied in machine learning, mathematics, physics, and other domain sciences \cite{Kolaczyk2009, DBLP:journals/corr/BrugereGB16}. Often the entities and relationships are unambiguously known: two users are `friends' in a social network, or two road segments are adjacent in the network if they physically intersect. Often, an underlying social, biological or other process generates data with latent relationships among entities. Rather than studying the process of interest through either coarse population-level statistics or isolated individual-level statistics, networks tend to represent complexity at multiple scales, and are general and reusable representations for different questions of interest on the process generating the original data.  

Previous work often focuses on ad-hoc, rule-based network construction, or model-based representational learning. Our flexible, general framework encompasses and formalizes these approaches. Our framework also learns networks subject to many targets: cascade modeling and routing, node and edge classification, or influence in networks. 

\section{General Framework}

Our framework transforms data from individual entities that are unambiguously known (e.g. users, IP addresses, genes) represented as nodes, into a \emph{space} of networks which are then sampled under a set of cost constraints, and evaluated relative to a problem of interest.

For a set of nodes $\{v_i \in \mathbf{V}\}$, a set of edge weight probability density functions $\{\mathbf{d}_{ij}() \in \mathbf{D} : (i,j) \in (V \times V), \mathbf{d}_{ij}() \sim [0,1]\}$, a node attribute set $\{a_i \in \mathbf{A}\}$, and a node label set $\{l_i \in \mathbf{L}\}$  let $\mathbb{G} = (\mathbf{V}, \mathbf{D}, \mathbf{A}, \mathbf{L})$ be a \emph{space} of weighted, attributed graphs.\footnote{Node labels are simply a specific \emph{node attribute} of interest for a subsequent task, defined separately for notational convenience} A weighted, attributed graph $\mathbf{G}'$ = $(\mathbf{V}, \mathbf{E}', \mathbf{A}, \mathbf{L})$ is drawn from $\mathbb{G}$ by sampling each edge weight distribution: $\mathbf{E'}= \{e'_{ij} \sim \mathbf{d}_{ij}() : (i,j) \in (V \times V)\}$. 

\begin{figure}[h]
\includegraphics[width=0.99\columnwidth]{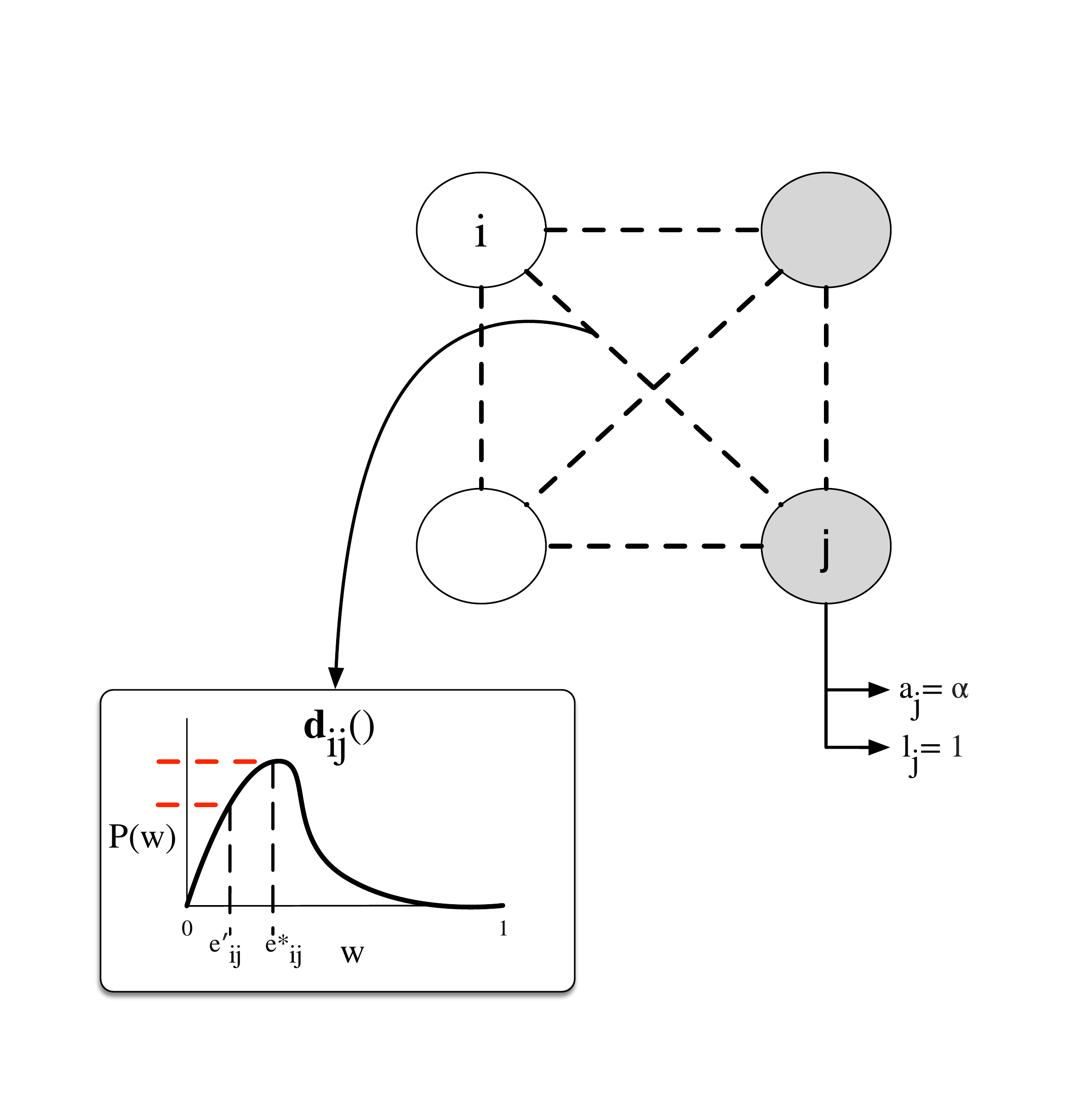}
\caption{A schematic of our framework, showing a network space for 4 nodes. Each node $v_i$ has a label value $l_i$ ($l_i$=1 shaded), and a linear threshold attribute $a_i$, set globally in this example. Each dashed edge between nodes denotes an edge weight density function. The shown $\mathbf{d}_{ij}()$ has an associated maximum likelihood, $e^*_{ij}$, an edge weight sample $e'_{ij}$, and the independent edge loss (red dashed lines).}
\label{fig-schematic}
\end{figure}

Our general framework evaluates weighted graphs within $\mathbb{G}$ according to some task $\mathcal{T}(\mathbf{G}', \bullet)$ subject to loss $\mathcal{L}_{\mathcal{T}}(G')$. See Figure \ref{fig-schematic} for a schematic of this formalization.

\section{Problem Formulation: Linear Threshold Process}

We instantiate a particular task on the above framework, to sample weighted networks which model a set of observed node labels $\mathbf{L}$ as the result of an \emph{Linear Threshold} spreading process \cite{Kempe:2003:MSI:956750.956769}.

Given a weighted network, $\mathbf{G}'$, the linear threshold process is initialized with $k$ labeled nodes. At each time step, unlabeled nodes adopt the label of the neighborhood if the sum of neighbor weights exceeds the node's threshold. The process continues until label assignments stabilize.\footnote{This assumes the susceptible-infected model. Nodes with $l_i=0$ were necessarily never infected over the process.} For simplicity, we'll instantiate a global threshold ($a_{i} = \alpha$, for all $a_i \in \mathbf{A}$) binary label formulation ($l_i \in \{0,1\}$ for all $\l_i \in \mathbf{L}$).

\begin{algorithm2e}
  \SetAlgorithmName{Problem}
  \\
  \\
  \KwGiven{Graph-space $\mathbb{G} = (\mathbf{V}, \mathbf{D}, \mathbf{A}, \mathbf{L})$ with $\mathbf{A}$ node linear thresholds}
  \KwFind{$\mathbf{E}' \sim \mathbf{D}$ and $k$ nodes: $S \subseteq \mathbf{V}$}
  \KwWhere{$\mathbf{L}$ is realized on $\mathbf{E}'$ through a linear threshold process initialized on seed-set $S$}
  \KwMin{$k$}
  \caption{Network Inference LT $k$-Seed Selection}
  \label{p:unconstrained}
\end{algorithm2e}

This problem aims to find the smallest set of nodes which produce the observed label set $\mathbf{L}$, under thresholds $\mathbf{A}$, when initializing the Linear Threshold model on the selected nodes $S$. A trivial solution exists where $k=|l_i = 1|$, the total number of labeled nodes. We construct this solution by setting edges incident to $l_i = 0$ to 0 weight, and the final $\mathbf{L}$ is trivially realized on initialization.

In the above problem, we are unconstrained by any loss function $\mathcal{L}(\mathbf{G}')$. Therefore we are always ensured at least $|l_i = 1|$ minimal solutions, with $k=1$. 

\emph{Proof sketch}: Selecting any node where $l_i = 1$, we set edges incident to $v_i$ where $l_j = 1$ such that $a_j$ is satisfied (simply: $e_{ij}=1$), therefore we infer a star with binary edges of all labeled nodes for each solution.

We allow our method to adjust $e'_{ij}$ directly rather than sampling randomly from $\mathbf{d}_{ij}()$. Recall that the range of $\mathbf{d}_{ij}()$ is $[0,1]$. Therefore even if $\mathtt{P}(\mathbf{d}_{ij}()=0)=0$, we allow setting $e'_{ij}=0$. In the constrained case (below) we will be penalized for this unlikely or unobserved edge weight.

\subsection{Independent Edge Loss}

We introduce a loss function measuring edge density function likelihood. This will incur cost when setting edge weights $E'$, conditioned on the respective edge density function. 

The independent edge loss measures the likelihood of a sampled edge, $e'_{ij}$ against the edge's maximum likelihood estimate: $e^*_{ij} = \mathtt{MLE}({d}_{ij}())$:

\begin{equation}
	\mathcal{L}(e'_{ij}, e^*_{ij}) = \mathtt{P}(\mathbf{d}_{ij}()=e^*_{ij}) - \mathtt{P}(\mathbf{d}_{ij}()=e'_{ij})
\end{equation}

Defined over an entire realized graph, we get:

\begin{equation}
	\mathcal{L}(\mathbf{G}') = \displaystyle \sum_{(i,j) \in (\mathbf{V} \times \mathbf{V})}\mathcal{L}(e'_{ij}, e^*_{ij})
\end{equation}

\begin{algorithm2e}
  \SetAlgorithmName{Problem}
  \\
  \\
  \KwGiven{Graph-space $\mathbb{G} = (\mathbf{V}, \mathbf{D}, \mathbf{A}, \mathbf{L})$ with $\mathbf{A}$ node linear thresholds, budget $\lambda$}
  \KwFind{$\mathbf{E}' \sim \mathbf{D}$ and $k$ nodes: $S \subseteq \mathbf{V}$}
  \KwWhere{$\mathbf{L}$ is realized on $\mathbf{E}'$ through a linear threshold process initialized on seed-set $S$}
  \KwMin{$k$, subject to $\mathcal{L}(\mathbf{G}') \leq \lambda$}
  \caption{Budgeted Network Inference LT $k$-Seed Selection}
  \label{p:constrained}
\end{algorithm2e}

Problem \ref{p:constrained} adds the Independent Edge Loss constraint to the initial Network Inference LT $k$-Seed Selection problem, also accepting as input a loss budget $\lambda$. 

\subsection{Existence of solutions}

Problem \ref{p:constrained} under infinite budget $\lambda = \infty$ is equivalent to Problem \ref{p:unconstrained}, yielding the same $k=1$ solutions. Depending on finite $\lambda$, we \emph{cannot} guarantee the existence of a solution.  When $\lambda=0$, there exists exactly one potential solution, the maximum likelihood edge weight set $\mathbf{E}^*$, which may not produce $\mathbf{L}$ under any seeding. 

\emph{Proof, by example}: Let $\mathbf{E}^*$ be a star with binary edge weights: $\mathbf{E}^* = \{e'_{ij} \in \{0,1\}\}$. Let the center of the star, $v_i$ be unlabeled: $l_i=0$. The periphery of the star, $S = \mathbf{V} \ v_i$ are all labeled: $l_j=1$. S must therefore be the seeds of the linear process. This is because $v_i$ is unlabeled therefore will not propagate to any $v_j \in S$. Because all $v_j \in S$ are labeled, the binary edge weights incident to $v_i$ satisfy \emph{any} threshold $a_i \in [0,1]$, so $v_i$ must be labeled after the linear threshold process: $l_i = 1$. Therefore $\mathbf{L}$ is not realizable on $\mathbf{E}^*$

\subsection{First Approximation}

Rather than formulating a solution to the weighted network case, let's consider only the binary case. In this case, we realize a network which satisfies our loss budget $\mathcal{L}(\mathbf{G}') \leq \lambda$, where $e'_{ij} \in \{0,1\}$, 

The Influence Maximization $k$-seed selection problem generates candidates for the Budgeted Network Inference LT $k$-Seed Selection problem, under this added constraint. A selected $k$ seed labels cannot propagate through unlabeled $v_i$. Therefore we assume edges incident to unlabeled $v_i$ will not satisfy $a_i$. This effectively disconnects each $v_i$, yielding connected components of nodes labeled in $\mathcal{L}$.  An accepted seed-set for each sub-problem is the Influence Maximization $k$-seed selection solution which labels all nodes in the component. The total seed set is the union of these sub-problem seed sets.

This approximates the optimal $k$ for one particular $\mathbf{G}'$ realization. However, it remains an open problem exploring the graph-space in an efficient way to improve a particular $\mathbf{G}'$ realization with respect to $k$.

\section{Other Formulations}

This general pattern of $\lambda$-constrained graph sampling in $\mathbb{G}$, subject to Independent Edge Loss generalizes to diverse tasks. For example, collective classification \cite{sen:aimag08}, is instantiated on labels $\mathbf{L}$ used to train local classifiers on node attributes $\mathbf{A}$. We sample $\mathbf{G}'$ on this task to maximizes classifier performance under $\lambda$ loss constraints.

In the area of influence maximization and information networks, this framework can also incorporate different transmission models (e.g. independent cascade), as well as parameterized rates of transmission on edges \cite{GomezRodriguez:2010:IND:1835804.1835933}. In information networks, edge weight density can be empirically measured from \emph{delay times} between information arrival at nodes. Once again, $\lambda$-constrained graph sampling in $\mathbb{G}$ realize graphs to predict known cascades of information, which may perform better for prediction than the MLE graph $\mathbf{G}*$. 

\section{Conclusion and Open Problems}

This is only a brief outline of this general network inference framework for modeling non-network data for a particular task. We sample a space of networks from observed data, subject to loss constraints and a task objective. Future work will focus on efficient search strategies which take advantage of shared problems across tasks, and comparing graph-edit heuristics across different tasks.

% labeled nodes
%
%  on $\mathbf{E}'$ If  is a valid solution For a valid solution, we observe that labels cannot propagate through , therefore   uses the Dominating Set

% \subsection{Monotonicity of $k$ vs. thresholds}
%
% Although we set all $a_i= \alpha$ for simplicity, this assumption was not exploited in our problem formulation. $k$ is monotonically increasing with the increase of any $a_i$. Consider the above binary-edge case. If we increase an $a_i$ on a node with final label $l_i=1$, causing the node to be unlabeled at the end of the process, we must add one or more seeds under any current seed set to correctly label $v_i$.\footnote{Note that } Similarly, if we increase $a_i$ to a node with final label $l_i=0$ and it is correctly labeled the process initialized on our seed set, it will remain correctly labeled $l_i=0$ under the larger $a_i$.

%According to L  One trivial instance generates $\mathbf{E}^* =\{e_{ij}=0\}$

%We can find a better $k$ by observing that a label cannot propagate through a node where $l_i=0$ at the end of the process. Therefore, we can decompose the network into labeled connected components by setting edge weights incident to nodes $l_i = 0$ to 0.   

%\section{Related Work}

%Inferring information networks: \cite{Gomez-Rodriguez:2012:IND:2086737.2086741, ICML2011Gomez_354}

%Task-oriented: \cite{McAuley:2015:INS:2783258.2783381, DeChoudhury:2010:IRS:1772690.1772722, Li2015}

%\subsection{Analysis}

%\section{Future Work}
  
\bibliographystyle{plainnat}
\bibliography{ivan}

\end{document}